\documentclass[fianl,1p,times]{elsart}

\usepackage{graphicx}
\usepackage{amssymb}

\usepackage{pseudocode}
\journal{Physics A}
\begin{document}

\begin{frontmatter}
\title{Information Filtering via Collaborative User Clustering Modeling}

\author[inst1]{Chu-Xu Zhang}
\author[inst1]{,Zi-Ke Zhang}
\corauth{zhangzike@gmail.com}
\author[inst1]{,Lu Yu}
\author[inst1]{,Chuang Liu,}
\author[inst1]{Hao Liu,}
\author[inst2]{Xiao-Yong Yan}
\corauth{yanxy@mail.bnu.edu.cn}

\address[inst1]{Institute of Information Economy, Hangzhou Normal University, Hangzhou 310036, PRC}
\address[inst2]{School of Systems Science, Beijing Normal University, Beijing 100875,PRC}

\begin{abstract}
The past few years have witnessed the great success of recommender systems, which can significantly help users find out personalized items for them from the information era. One of the most widely applied recommendation methods is the Matrix Factorization (MF). However, most of researches on this topic have focused on mining the direct relationships between users and items. In this paper, we optimize the standard MF by integrating the user clustering regularization term. Our model considers not only the user-item rating information, but also takes into account the user interest. We compared the proposed model with three typical other methods: User-Mean (UM), Item-Mean (IM) and standard MF. Experimental results on a real-world dataset, \emph{MovieLens}, show that our method performs much better than other three methods in the accuracy of recommendation.
\end{abstract}

\end{frontmatter}

\section{Introduction}
Facing the explosive growth of web, people would become lost in the web's info-thickets and often waste much time to search useful and personalized information, namely $Information\ Overload$. Confronting with such problem, researchers from different areas have invented various tools, among which search engine is the most outstanding. However, compared with recommender systems\cite{resnick1997recommender,lu2012recommender}, which automatically match the user's taste based on the historical behaviors, search engine is not personalized enough because it produces the same result for all users. Among kinds of recommender systems, Collaborative Filtering (CF) \cite{herlocker2000explaining,schafer2007collaborative} is the most widely used in different fields due to its advantages of requiring no domain knowledge, implementing easily and detecting the complex pattern that is hard to be exploited with the known data. As a result, CF has attracted much attention from both academic and industry fields in the past decade. In particular, the competition of Netflix Prize (NP) \cite{bennett2007netflix}, has inspired different fields of  researchers to propose various solutions to build corresponding recommender systems.

The basic idea of CF is that recommendation for the target user is made by predicting the preference of the uncollected items based on the neighbors. Neighbor is a group of persons with similar tastes when they rate the same items. Generally, there are two main types of CF: neighborhood
and model based approaches \cite{su2009survey}. In the early time, neighborhood based methods, including user and item based approaches, were the the most widely applied in the industry, such as Amazon \cite{linden2003amazon}, Google \cite{liu2010personalized}. In recent years, experts from both academic and industry have witnessed the excellent performance of model-based approaches, especially the Latent Factor Model (LFM) \cite{koren2009matrix}. As the typical representative technique of LFM based methods, matrix factorization (MF) provides an alternative method to represent the relationship between users and items. In the LFM, users and items are both represented in the same latent factor space (LFS), hence the prediction is accomplished by directly evaluating the preferences of users for the uncollected items. Some MF methods \cite{rennie2005fast,salakhutdinov2008bayesian,salakhutdinov2008probabilistic,srebro2003weighted} have been proposed
in CF because of the high efficiency in dealing with large-scale data sets. Those approaches tend to fit the
user-item rating matrix with low-rank matrix factorization and apply it to make rating predictions. MF is efficient in training since it assumes that only few factors influence preferences in user-item ratings. The objective for minimizing the sum-squared errors can be easily solved by Singular Value Decomposition (SVD), and Expectation Maximization(EM) algorithms for solving weighted low-rank approximation was proposed in \cite{srebro2003weighted}.

Since the success of MF in the Netflix Prize competition, a great many of variants are proposed. In \cite{ma2011recommender}, a matrix factorization framework with social network regularization was described. It provided a general method for improving recommender system by incorporating social network information. Ma $et\ al.$ \cite{karatzoglou2011collaborative} presented two simple models that take advantages of the temporal order of choices and ratings. These two models not only exploited the collaborative effects in the data, but also took into account the order in which items could be viewed by the users. Koren $et\ al.$ \cite{rendle2010factorizing} introduced a Markov Chain model which considered the collaborative effects using Tensor Factorization \cite{welling2001positive}.

In addition, besides the traditional CF methods in recommender system, there also emerged many variant methods based on statistical physics with the development of network science, such as \cite{zhou2007bipartite,sun2009information,zhang2010personalized,zhou2010solving,shang2010collaborative,lu2011information,liu2012solving}. Some of these methods are innovative and effective in improving not only recommender accuracy but also recommender diversity and novelty. Zhang $et\ al.$ \cite{zhang2010personalized} proposed a recommendation algorithm based on an integrated diffusion on user-item-tag tripartite graphs and significantly improved accuracy, diversification and novelty of recommendations. In \cite{zhou2010solving}, a new algorithm specifically addressed the challenge of diversity in recommender system is proposed. L\"{u} $et\ al.$ \cite{lu2011information} introduced a recommendation algorithm based on the preferential diffusion process on user-object bipartite network.

In this paper, inspired by SNMF \cite{ma2011recommender} and tripartite network with tag system \cite{zhang2010personalized}, we consider the neighbors' impact on the interest of each user in the same LFS and propose a
recommendation model based on clustering users (UCMF). Firstly, we represent the interest of each user with the statistical information of her behaviors on different tags. Secondly, we classify all users into several groups by the K-Means clustering algorithm. Finally, we expand the standard MF by integrating a user clustering regularization term which describes that the users in the same group have similar interest. The results of an empirical analysis on \emph{MovieLens} show that our model outperforms the standard MF method and other two baseline methods in the accuracy of recommendation.
\section{User Clustering Model}
\subsection{Low-Rank Matrix Factorization}
CF techniques based on MF method assume that users' ratings on items can be represented by a $N\times M$ matrix ($N$ is the number of users and $M$ the number of items). A low-rank matrix factorization approach tends to approximate the rating matrix $R$ by  multiplying $L$-rank factors,
\begin{equation}
\label{EQ:eq1}
\ R\approx U^{T}V,
\end{equation}
where $U\in \mathbb{R}^{L\times N}$ and $V\in \mathbb{R}^{L\times M}$ with $L<min(N,M)$, the matrix $R$ is usually quite sparse.

Traditionally, the SVD method is employed to approximate the rating matrix $R$ by minimizing
\begin{equation}
\label{EQ:eq2}
\frac {1} {2}||R-U^{T}V\parallel _{F}^{2},
\end{equation}
where $\parallel \cdot \parallel$ denotes the Frobenius form. We only need to factorize the observed ratings in matrix $R$ because of large missing values. So Equation (\ref{EQ:eq2}) can be transformed to
\begin{equation}
\label{EQ:eq3}
\min _{U,V}\frac {1} {2}\sum _{i=1}^{N}\sum _{j=1}^{M}I_{ij}\left( R_{ij}-U_{i}^{T}V_{j}\right) ^{2},
\end{equation}
where $I_{ij}$ equals to 1 if user $u_{i}$ rates item $v_{j}$ and 0 otherwise.

Two regularization terms are added into loss function in order to avoid over-fitting. Hence the objective function becomes
\begin{equation}
\label{EQ:eq4}
\min _{U,V}\frac {1} {2}\sum _{i=1}^{N}\sum _{j=1}^{M}I_{ij}\left( R_{ij}-U_{i}^{T}V_{j}\right) ^{2}+\frac {\lambda _{1}} {2}\left\| U\right\| _{F}^{2}
+\frac {\lambda _{2}} {2}\left\| V\right\| _{F}^{2},
\end{equation}
where $\lambda _{1},\lambda _{2}>0$. The optimization problem in Equation \ref{EQ:eq4} is to minimize the sum-of-squared-errors objective function with quadratic regularization terms.
\subsection{Model}
%\subsubsection{K-Means Clustering Algorithm}
%In order to divide user into different interest clusters, we review K-Means clustering Algorithm first.
%In the clustering problem, we divide the data set $\left\{ x^{\left( 1\right) },\ldots ,x^{\left( m\right) }\right\}$ into a few cohesive $clusters$.
%It is an unsupervised machine learning problem since no labels $y^{(i)}$ of $x^{(i)}$ are given.
%The K-Means clustering algorithm is as follows:
%\begin{itemize}
%  \item Initialize cluster centers $u_{1},u_{2},\ldots ,u_{k}\in R^{n}$ randomly.
%  \item Repeat until convergence:{
%
%  For every $i$, set $C^{\left( i\right) }:=\arg \min _{j}\left\| x^{\left( i\right) }-u_{j}\right\| ^{2}$.
%
%  For every $j$, set $u_{j}:=\frac{\sum _{i=1}^{m}1\left\{ c^{\left( i\right) }=j\right\} x^{\left( i\right) }}{\sum _{i=1}^{m}1\left\{ c^{\left( i\right) }=j\right\}}$
%
%  }
%\end{itemize}
%In the above algorithm, $k$ is the number of clusters we want to find and the cluster centers $\mu^{j}$ represent current guess for
%the positions of the centers of the clusters. we set $k$ cluster centers values randomly for initialisation. The inner-loop of the algorithm
%repeatedly carries out two steps:$(i)$ Set each example $x^{(i)}$ to the closet cluster centroid $\mu_{j}$. $(ii)$ Move each cluster
%centroid $\mu_{j}$ to the mean of the points assigned to it. K-Means algorithm converge to local minimum until enough iterations.
\subsubsection{User Clustering}
\begin{figure}[htb]
\centering
  \includegraphics[width=6cm,height=6cm]{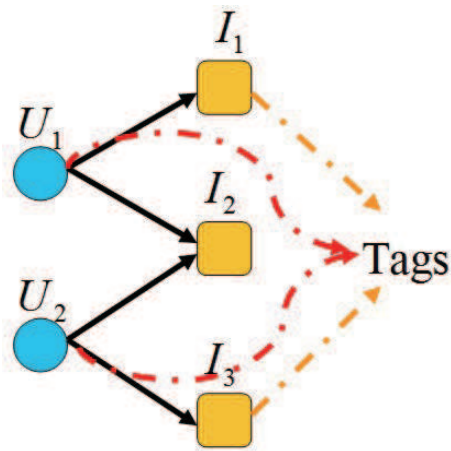}
  \caption{\label{Fig:fig1} Illustration of users' behavior on different items with the same tags.
  In one way, user $U_{1}$ and $U_{2}$ choose the same item $I_{2}$, which may indicate $U_{1}$ and $U_{2}$ have similar tastes.
  In the other way, $U_{1}$ and $U_{2}$ also choose different items $I_{1}$ and $I_{3}$ which have the same tags. This also means these two
  users should have similar interests in some extent.}
\end{figure}

Regular MF methods focus on the user-item ratings problem. However, they might miss some information which may help to improve the performance of
recommender system. In order to overcome this weakness, we cluster users by K-Means algorithm based on the statistics of user behaviors on different tags. As shown in Figure \ref{Fig:fig1}, different users may choose different items which have similar properties labeled by tags and those users tend to have similar interest. Inspired
by this general phenomenon in users' behaviors, we assign user $i$ a $H$-dimension interest vector ($H$ means the overall number of different tags)
\begin{equation}
\label{EQ:eq5}
\begin{array}{rcl}
 T_{i}&=&(t_{i1},t_{i2},\ldots,t_{iH}),\\
 t_{ih}&=&{\displaystyle\sum _{j=1}^{M}I_{ij}\cdot \delta_{jh}},
 \end{array}
\end{equation}
where $\delta_{jh}$ equals to 1 when tag $h$ is assigned to item $j$ and vice versa.
The $N\times H$ interest matrix $T$ records the cumulative statistics of each user' behavior on each tag.
We normalize $T$ to $\hat{T}$  %due to the fact that we focus on the preference on different tags% %of each user
and denote $\mu_{p}$ as the center of cluster $p$.
%and initialize it by choosing randomly a tag vector from \hat{T}.
$C$ is a $N$-dimension row vector and $C_{i}$ means the cluster that user $i$ is assigned to. $\theta_{0}$ guarantees the algorithm to be converged and $K$ is the overall number of user clusters. $B(C_{i}=j)$ equals to 1, when $C_{i}=j$ and 0 otherwise.
Our user clustering algorithm is described in algorithm \ref{Algorithm:Algorithm 1}.

\begin{pseudocode}[ruled]{User Clustering}{\hat{T},C}
\label{Algorithm:Algorithm 1}
\COMMENT{ Classify each user to different clusters by K-Means}\\
 \hat{T}\in \mathbb{R}^{N\times H},\ %\GETS (\hat{t_{i1}},\hat{t_{i2}},\ldots,\hat{t_{iH}})%,
 \mu\in \mathbb{R}^{K\times H}, \mu_{p} \GETS a\ random\ row\ from(\hat{T})\\
 C_{i}=\arg \min _{p}\left| \left| \hat{T_{i}}-\mu_{p}\right| \right| ^{2},J \GETS \sum _{i=1}^{N}\left\| \hat{T_{i}}-\mu_{C_{i}}\right\| ^{2} \\
\WHILE J > \theta_{0} \DO
\BEGIN
\FOR  i \GETS 0 \TO N-1 \DO C_{i} \GETS \arg \min _{p}\left| \left| \hat{T_{i}}-\mu_{p}\right| \right| ^{2}\\
\FOR  p \GETS 0 \TO K-1 \DO \mu_{p}\GETS \frac {\sum _{i=1}^{N}B\left( C_{i}=j\right) \hat{T_{i}}} {\sum _{i=1}^{N}B\left( C_{i}=j\right) }\\
J \GETS \sum _{i=1}^{N}\left\| \hat{T_{i}}-\mu_{C_{i}}\right\| ^{2}\\
\END\\
\RETURN{C}
\end{pseudocode}
\subsubsection{Regularization and Optimization}
Subsequently, we propose a collaborative user clustering model based on matrix factorization technique,
\begin{equation}
\label{EQ:eq6}
\begin{array}{rcl}
\min _{U,V}L(R,U,V&)&={\displaystyle \frac {1} {2}\sum _{i=1}^{N}\sum _{j=1}^{M}I_{ij}\left( R_{ij}-U_{i}^{T}V_{j}\right) ^{2}}
\\&&{\displaystyle
 +\frac {\alpha } {2}\sum _{i=1}^{N}\sum _{f\in G\left( i\right) }Sim\left( i,f\right) \parallel U_{i}-U_{f}\parallel _{F}^{2}}
\\&&{\displaystyle+\frac {\lambda _{1}} {2}\left\| U\right\| _{F}^{2}+\frac {\lambda _{2}} {2}\left\| V\right\| _{F}^{2}},
\end{array}
\end{equation}
where $\alpha>0$ and $G(i)$ is the set of users who are in the same cluster with user $U_{i}$, called cluster neighbors. $Sim(i,f)\in [0,1]$ is the similarity function to indicate the similarity between user $U_{i}$ and $U_{f}$. In this objective function, we incorporate user clustering regularization term to impose constrain between one user and their cluster neighbors individually,
\begin{equation}
\label{EQ:eq7}
\frac {\alpha } {2}\sum _{i=1}^{N}\sum _{f\in G\left( i\right) }Sim\left( i,f\right) \parallel U_{i}-U_{f}\parallel _{F}^{2}.
\end{equation}
The above clustering regularization term assumes that every user's interest is close to tastes of this user's cluster neighbors. A small value of $Sim(i,f)$ indicates that the distance between feature vector $U_{i}$ and $U_{f}$ should be larger, while large value tells that the distance between the feature vectors should be smaller. It is sensitive to those users whose cluster neighbors have diverse tastes, which may be important in accurately modeling the users' feature vectors.

Optimization solution of the objective function given by the equation \ref{EQ:eq6} can be found by performing Stochastic Gradient Descent (SGD) \cite{koren2009matrix} in the latent feature vector $U_{i}$ and $V_{j}$,
\begin{equation}
\label{EQ:eq8}
\begin{array}{rcl}
\frac {\partial L}{\partial U_{i}}&&=\displaystyle\sum _{j=1}^{M}I_{ij}\left( U_{i}^{T}V_{j}-R_{ij}\right) V_{j}+\lambda _{1}U_{i}
\\&&{\displaystyle+\alpha \sum _{f\in G\left( i\right) }Sim\left( i,f\right) \left( U_{i}-U_{f}\right)}
\\&&{\displaystyle-\alpha \sum _{f\in G\left( i\right) }Sim\left( i,f\right) \left( U_{f}-U_{i}\right)},
\\\frac {\partial L} {\partial V_{j}}&&{{\displaystyle=\sum _{i=1}^{N}I_{ij}\left(U_{i}^{T}V_{j}-R_{ij}\right) U_{i}+\lambda _{2}V_{j}}}.
\end{array}
\end{equation}
%we can use SGD to optimize the objective function \ref{EQ:eq7}.

We denote $\hat{R}$ the prediction result of rating matrix and $R_{train}$ the training set of our model.
The solution is described in algorithm \ref{Algorithm:Algorithm 2}.

\begin{pseudocode}[ruled]{Collaborative User Clustering Modeling}{U,V,R}
\label{Algorithm:Algorithm 2}
\COMMENT{Optimize the latent features of users and items}\\
 U\in \mathbb{R}^{L\times N},\ %\GETS (u_{i1},u_{i2},\ldots,u_{iL})%
 V\in \mathbb{R}^{L\times M}\ and\ with\ small\ random\ values\\ %\GETS (v_{j1},v_{j2},\ldots,v_{jL})%
 \WHILE R_{ij}\ in\ observations\ R_{train} \DO
\BEGIN
 U_{i}\GETS U_{i}-\eta\frac {\partial L} {\partial U_{i}}\\
 V_{j}\GETS V_{j}-\eta\frac {\partial L} {\partial V_{j}}\\
\hat{R} \GETS U^{T}V\\
\END\\
\RETURN{\hat{R}}
\end{pseudocode}
\subsubsection{Similarity Function}
In the above section, the proposed user clustering regularization term requires the knowledge of similarities between users. Since we have the rating information of all users, the evaluation of similarities between two users can be calculated by measuring the common ratings of these
two users $i$ and $j$. One of the most popular methods is Vector Space Similarity (VSS) \cite{breese1998empirical}. VSS is employed to define the similarity between two users and based on the items they rated in common,
\begin{equation}
\label{EQ:eq9}
Sim(i,f)=\frac{\sum _{j\in I\left( i\right) \cap I\left( f\right) }R_{ij}R_{fj}}{\sqrt {\sum _{j\in I\left( i\right) \cap I\left( f\right) }
R_{ij}^{2}}\cdot \sqrt {\sum _{j\in I\left( i\right) \cap I\left( f\right) }R_{fj}^{2}}},
\end{equation}
where $j$ belongs to the subset of items which both user $i$ and $f$ have rated. $R_{ij}$ means the rate user $i$ gives to item $j$.
We can see that $Sim(i,f)$ is within the range [0,1] and a larger value means user $i$ and $f$ are more similar with each other.
\section{Experiment}
\subsection{Experiment Setup}
\subsubsection{Data}
We evaluate the proposed model by a benchmark dataset \emph{MovieLens}, which consists of approximate one million(1M) ratings by 6,040 users and 3,952 movies. Each movie is rated in a scale from 1 to 5 stars and labeled with several tags like comedy, romance and so on.
\subsubsection{Evaluation Protocol}
We use two metrics, the Mean Absolute Error (MAE) and the Root Mean Square Error (RMSE), to measure the prediction quality of our proposed
approach in comparison with other methods. The metric MAE is defined as:
\begin{equation}
\label{EQ:eq10}
MAE=\frac {1} {S}\sum _{i,j}\left| R_{ij}-\widehat {R}_{i,j}\right|,
\end{equation}
where $R_{ij}$ denotes the rating user $i$ gave to item $j$, $\widehat {R}_{ij}$ denotes the rating user $i$ gave to item $j$ as predicted by a method, and $S$ denotes the number of tested ratings. The metric RMSE is defined as:
\begin{equation}
\label{EQ:eq11}
RMSE=\sqrt {\frac {1} {S}\sum _{i,j}\left( R_{ij}-\widehat {R}_{ij}\right) ^{2}}.
\end{equation}
Obviously, the smaller MAE or RMSE is, the better performance the algorithm will be.
\subsection{Results}
\subsubsection{Comparisons}
We conduct experiments to assess the performance of our model.
In addition, we compare our recommendation results with the following methods:
\begin{itemize}
  \item User-Mean (UM): this method uses the mean value of all users to predict the missing values.
  \item Item-Mean (IM): this method utilizes the mean value of all items to predict the missing values.
  \item MF: this is the regular MF method and it is widely used in collaborative filtering recently. It only uses user-item rating matrix for
  recommendations.
\end{itemize}
For the \emph{MovieLens} dataset, we use different training data settings (90\%,80\%,70\%) to test the algorithm.
The random selection is carried out 10 times independently, as we report the average results. The standard deviation of the results is
less than 0.001. We set $\lambda _{1}=\lambda _{2}=0.01$ and $\alpha=0.001$,. The size of latent feature $L$ and user cluster number $K$ are set to 10 and 5, respectively. The detailed comparisons are shown in Table 1.

From these results, we can observe that our method outperforms
other approaches in all the settings of this dataset. Note that we focus on the relative improvement of our model over MF.
UCMF outperforms the MF around 6\% both in MAE and RMSE. Our method not only considers the user-item rating information but also takes into account the users' interests, both of which have heavy impacts on the results of prediction.
\begin{table}[htbp]
\label{Tab:table 1}
\centering
\caption{Performance comparisons of four methods on \emph{MovieLens}. All of these results are obtained by
averaging over 10 runs, each of which has three independently random divisions (90\%,80\%,70\%) of training set.}
\begin{tabular}{ccccccccc}  \hline Trainning & Metrics & UM & IM & MF & $\mathbf{UCMF}$
\\ \hline
  90\% & MAE  & 0.91488 & 1.01714 & 0.79749 & $\mathbf{0.74876}$\\
  90\% & RMSE & 1.12326 & 1.24578 & 1.03255 & $\mathbf{0.94836}$\\
  80\% & MAE  & 0.91546 & 1.01860 & 0.80442 & $\mathbf{0.74947}$\\
  80\% & RMSE & 1.12388 & 1.25277 & 1.05058 & $\mathbf{0.95020}$\\
  70\% & MAE  & 0.91604 & 1.02038 & 0.80566 & $\mathbf{0.75093}$\\
  70\% & RMSE & 1.12397 & 1.25650 & 1.05625 & $\mathbf{0.95121}$\\
\hline
\end{tabular}
\end{table}
\subsubsection{Impact of parameter $\alpha$}
\begin{figure}[htb]
\centering
  \includegraphics[width=15cm,height=5cm]{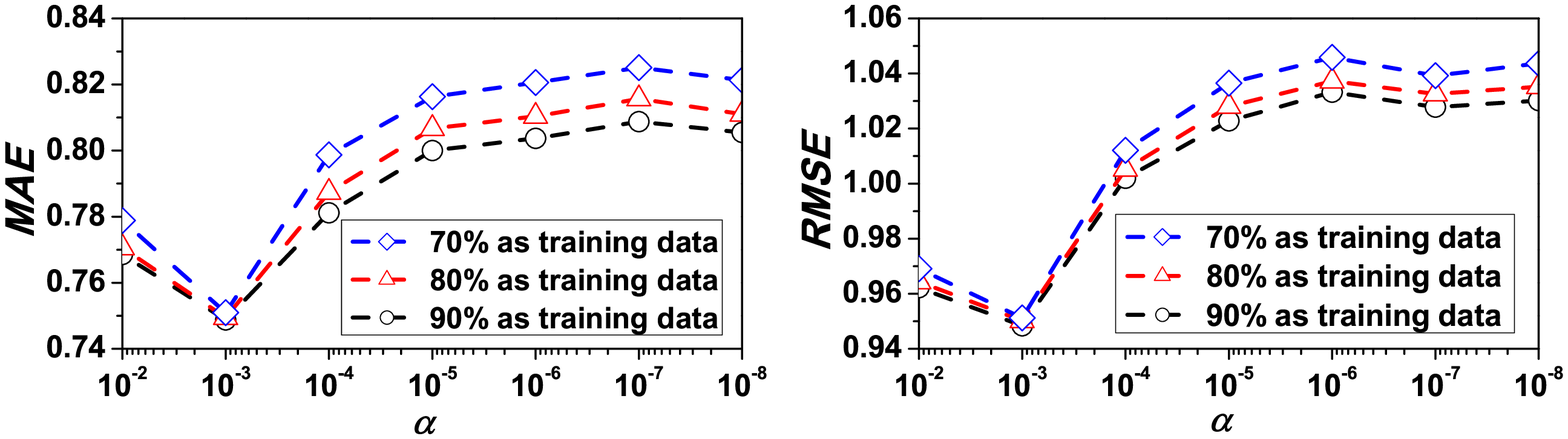}
  \caption{\label{Fig:fig2} Impact of parameter $\alpha$}
\end{figure}

In our proposed model, the parameter $\alpha$ plays an important role. It determines how much our method should incorporate user clustering information. In the extreme case, if we use a very small value of $\alpha$, we only use user-item rating matrix for MF, which simply employs users＊ own interests in making recommendation. On the other side, if $\alpha$ is very large, the information of user clustering may dominate the recommendation. In the following, we analyze how the changes of $\alpha$ can affect the recommendation accuracy. Fig. \ref{Fig:fig2} shows the impact of $\alpha$ on MAE and RMSE in our model. We can find the significant impact of value $\alpha$ on recommendation results, which means the effect of user clustering information is great. From the results, as $\alpha$ increases, the MAE and RMSE values decrease at first, but when $\alpha$ goes below a certain threshold $(\alpha=0.001)$, the MAE and RMSE values increase with further decrease of $\alpha$. The existence of certain point of $\alpha$ confirms that the appropriate integration of user-item rating matrix and user clustering information can result in the optimal recommendation.
\subsubsection{Impact of user clustering number $K$}
\begin{figure}[htb]
\centering
  \includegraphics[width=15cm,height=5cm]{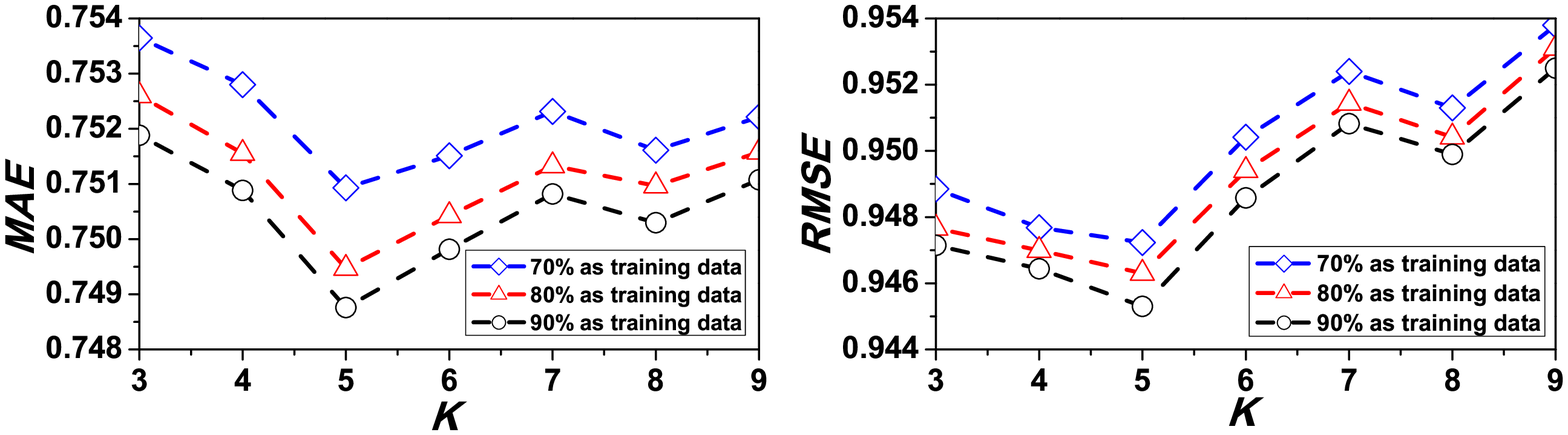}
  \caption{\label{Fig:fig3} Impact of user clustering number $K$}
\end{figure}

The number of user clusters may make great impact on accuracy of recommendation. Each user may belong to different clusters due to different
clustering number. In order to examine how much the clustering number $K$ impact on the results of whole model, we also conduct an experimental
analysis on different clustering number settings. Fig. \ref{Fig:fig3} reports the relationship between metrics and $K$. With the increment of $K$, MAE and RMSE decrease first, but perform an increment trend when $K$ goes beyond a certain threshold value 5. This observation demonstrates the importance of appropriate user clustering number.

\section{Conclusions}
In this paper, we have presented an intuitive way of using the information of user behavior to cluster users and improve the quality of the
recommendation. The UCMF exploits more information about users＊ behaviors and interests than regular MF method. The statistics of
users' behaviors on different tags perform well in measuring the users＊ interests and clustering them.

Note that we take advantage of user clustering information, our model performs better than regular MF method and other two algorithms. We find that appropriate values of parameter $\alpha$ and user clustering number $K$ can make great impact on the accuracy of recommendation results. We also believe that this collaborative user clustering method will also be strong in other recommendation domains and plan to test this in the future work.
\section*{Acknowledgments}
This work was partially supported by the National Natural Science Foundation of China (Grant Nos. 11105024, 11105040 and 1147015), and the Zhejiang Provincial Natural Science Foundation of China (Grant Nos. LY12A05003 and LQ13F030015), the start-up foundation of Hangzhou Normal University. ZKZ acknowledges the EU FP7 Grant 611272 (project GROWTHCOM).
\bibliographystyle{unsrt}
\bibliography{v4}
%\end{thebibliography}
\end{document}